\documentclass[10pt,floatfix,showpacs,twocolumn]{revtex4}
\usepackage{amsmath}
\usepackage{amssymb}
\usepackage{bm}
\usepackage{epsfig}
\usepackage{graphicx}
\usepackage{color}

\renewcommand{\Re}{\mathop{\rm Re}\nolimits}
\renewcommand{\Im}{\mathop{\rm Im}\nolimits}

\newcommand{\rmi}{{\rm i}}
\begin{document}
\title{Collective modes of quantum dot ensembles in microcavities}

\author{N.S. Averkiev}
\author{M.M. Glazov}\email{glazov@coherent.ioffe.ru}
\author{A.N. Poddubny}

\affiliation{Ioffe Physical-Technical Institute RAS, 26 Polytekhnicheskaya, 194021 St.-Petersburg, Russia}

\begin{abstract}
Emission spectra of quantum dot arrays in zero-dimensional microcavities are studied theoretically, and it is shown that they are determined by the competition between the formation of the collective superradiant mode and inhomogeneous broadening. The random sources method for the calculation of photoluminescence spectra under a non-resonant pumping is developed, and a microscopic justification of the random sources method within a framework of the standard diagram technique is given. The emission spectra of a microcavity are analyzed with allowance for the spread of exciton states energies caused by an inhomogeneous distribution of quantum dots and a tunneling between them. It is demonstrated that in the case of a strong tunneling coupling the luminescence spectra are sensitive to the geometric positions of the dots, and the collective mode can, under certain conditions, be stabilized by the random tunnel junctions.
\end{abstract}

\date{\today}

\maketitle

\section{Introduction}
Cavity quantum electrodynamics is one of the most actual directions in the field of modern optics. It studies systems where a photon localized in one or several directions interacts with elementary excitations of a media~\cite{Dutra,Yamamoto}\:. Beams of atoms and single atoms and molecules can stand out for the media. Quantum properties of radiation and matter are most brightly manifest themselfs in these studies (see, e.g., \cite{schuster} and references therein).

Semiconductor nanostructures serve as the solid state analogs of such atomic systems. Investigation of the optical properties of semiconductor nanostructures, and, in particular, structures with quantum dots, so-called artificial atoms, is one of the rapidly developing areas of the modern solid state physics~\cite{ivchenko05a,bardot2005,scheibner2007}. The effective interaction between localized photon modes and excitons (electron-hole excitations) is attained in semiconductor quantum microcavities~\cite{kavokin03b}. Light-matter interaction being multiply enhanced by a  cavity allows to efficiently visualize eigenstates of nanostructures~\cite{averkiev07}.

Lately a remarkable progress in obtaining the strong coupling regime between an exciton localized in a single quantum dot  (QD) and a photon trapped in three dimensions by a  cavity is reached~\cite{reithmaier04a, yoshie04a, peter05a,
hennessy07}. In the case where the damping of a photon mode caused by the imperfections of a microcavity and a non-radiative damping of an exciton become smaller as compared to their coupling constant, new eigenstates of the system, zero-dimensional exciton polaritons, are formed. Eigenstates structure and emission spectra in such cavities have been studied theoretically in a number of works, see e.g.~\cite{andreani99a, laussy2006a,fabrice08}. It was assumed in these works that only one quantum dot is in the strong coupling regime, the possible presence of other emitters has been taken into account phenomenologically~\cite{fabrice08}.

In this regard, the study of optical properties of microcavities where several emitters with energies near the photon resonance are placed is of apparent interest. The present paper is aimed at the theoretical investigation of this problem. It can be expected, that the interaction with light will reinforce definite (symmetric or superradiant~\cite{scheibner2007,khitrova2007,Dicke54a})
modes in much the same way as it takes place in planar microcavities and in optical cavities with cold atoms~\cite{houdre96,kavokin03b}. However, as it will be shown here, the interaction of light with localized excitons in semiconductor quantum dots has a number of fundamental features being most brightly manifested in the presence of the tunneling coupling between the dots.

The paper structure is as follows. The model of zero-dimensional microcavity with a quantum dot array is formulated in Sec.~\ref{sec:model} and a formalism to calculate the photoluminescence (PL) spectra of such structures based on the random sources method is set forth. This method is microscopically justified within  the framework of the standard diagram technique for non-equilibrium systems. Third section of the paper is devoted to the investigation of different models of exciton generation in resonant and non-resonant quantum dots and related features in the luminescence spectra. The calculated emission spectra of the disordered quantum dot arrays with allowance for the dispersion of localized excitons resonance energies are given in the forth section. Photoluminescence of quantum dot arrays with tunneling coupling is investigated theoretically in Sec.~\ref{sec:tunnel}. Main results of the work are briefly discussed in Sec.~\ref{sec:concl}.

\section{Model}\label{sec:model}

Consider $N$ quantum dots, placed in a zero-dimensional microcavity. It is convenient to write the system Hamiltonian in the secondary quantization representation by introducing the operators $c$
($c^\dag$), describing the annihilation (creation) of a localized photon, and operators
$a_i$ ($a_i^\dag$), describing annihilation (creation) of excitons, localized in quantum dots ($i=1\ldots N$):
\begin{equation}\label{hamiltonian}\begin{split} 
\mathcal H = \omega_{\rm phot} c^\dag c + \sum_i \omega_i a_i^\dag a_i +& \sum_{ij}
t_{i,j} a_i^\dag a_j +\\& \sum_i \bigl(V_i c^\dag a_i + V_i^* c a_i^\dag\bigr).
\end{split}
\end{equation}
Here $\omega_{\rm phot}$ is the cavity photon mode energy, $\omega_i$ are the exciton energies in quantum dots, the Hermitian matrix $t_{i,j}$ ($t_{i,j}=t_{j,i}^*$) describes the tunneling coupling between the quantum dots, and constants $V_i$ describe an interaction of localized excitons with light. 

It is assumed that excitons tunnel between the dots as a whole because Coulomb interaction prevents charge separation~\cite{Lan00, nakaoka2006, bardot2005}\:. Such a situation can be realized in large quantum dots, whose typical scales are larger as compared to the exciton Bohr radius. Independent tunneling of an electron and a hole leads to an additional decay channel of the exciton resonance and can be taken into account by standard methods.

Let us assume for simplicity that the wavefunctions of exciton states can be chosen as real. Thus, by matching the signs of the wavefunctions, the quantities $V_i$ can be made positive. Note, that such a sign change determines the signs of the constants $t_{ij}$. Possible difference of the signs of the tunnelling constants and its effect on the optical spectra of quantum dots in a microcavity is discussed in Sec.~\ref{sec:tunnel}.

The cavity mirrors are not ideal, therefore photon mode turns out to be coupled with the continuum of states propagating outside of the cavity. On the one hand, it leads to a finite lifetime of a photon inside the cavity, and on the other, to the possibility to detect the emission of the system.

\begin{figure}
 \includegraphics[width=0.45\textwidth]{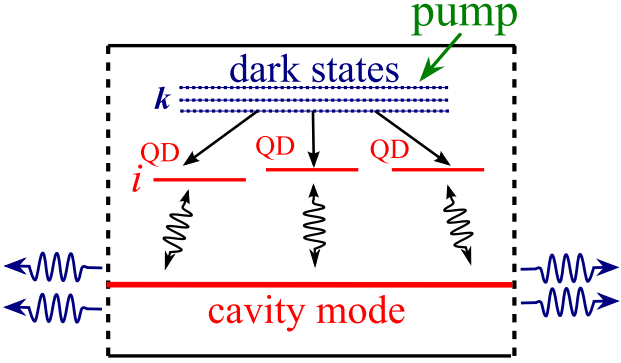}
\caption{Photoluminescence formation scheme in the microcavity with quantum dots (QD). Subscript $i$ denotes lowest in energy exciton states in dots. A non-resonant pumping (pump) to the excited  ``dark'' states, denoted by a multi-index $\bm k$, not coupled with the photon mode (cavity mode) is assumed. Radiation escapes microcavity due to the tunneling of a photon mode through the mirrors.}\label{fig:scheme}
\end{figure}

Schematic illustration of the photoluminescence formation in the considered system is given in Fig.~\ref{fig:scheme}. It is assumed that a non-resonant excitation takes place, i.e. the excitons or free electron-hole pairs are generated in the excited states whose energies are far from the photon mode. Pumping leads to the formation of the steady distribution function of carriers in the excited states in quantum dots. These states are ``dark'', because they are far in energy from the photon mode. The relaxation of carriers and excitons from the excited states to the lowest ones being in the resonance with photon mode takes place due to the interaction with phonons. The emission caused by the photon escape away from the cavity is detected.

\subsection{Random sources method}\label{sec:random_sources}

In order to calculate photoluminescence spectrum of quantum dot ensemble we have extended the random sources method developed in Refs.~\cite{VorIvch07,deych:075350} for the quantum well structures. It consists of the determination of the electric field induced as a result of exciton generation in quantum dots, characterized by the amplitudes of  the random sources $F_{i}$. Photoluminescence spectrum is determined by an intensity of the field in a cavity mode averaged over the realizations of random sources. Therefore, within  the framework of this method the inhomogeneous system of equations
\begin{align}
 (\omega_{\rm phot}-\omega-\rmi\gamma_{\rm phot})E&=\sum\limits_{i=1}^N
V_i P_{i}\label{motioneqs}\\
 (\omega_{i}-\omega-\rmi\gamma_{i})P_{i}+\sum\limits_{j\ne
i}t_{i,j}P_{j}&=V_iE+F_{i},\quad i=1\ldots N\:,\nonumber
 \end{align}
describing oscillations at a frequency $\omega$ of coupled oscillators, photon and excitons, under the action of driving forces $F_{i}$ is solved. Here, the quantity $E$ is proportional to the field amplitude in the photon mode of the microcavity, and the quantities 
$V_iP_{i}$ are proportional to the coordinate integrated contributions of excitons in 
$i$th quantum dot to the media polarization. Eqs.~\eqref{motioneqs} can be obtained from the Heisenberg equations of motion for the operators $c$, $a_j$, whose dynamics is governed by the Hamiltonian~\eqref{hamiltonian}, and non-radiative dampings of exciton
$\gamma_{i}$, photon damping $\gamma_{\rm phot}$ and amplitudes $F_i$ are added into Eq.~\eqref{motioneqs} in a phenomenological way. 

It is convenient to introduce the eigenfrequencies $\Omega_{m}$, determined from the homogeneous system 
(\ref{motioneqs}), as well as the eigenvectors
$[C^{(m)}_{\rm phot},C^{(m)}_{1}\ldots C^{(m)}_{1}]$ $\equiv$ $[E,P_1\ldots P_N]$,
which represent Hopfield coefficients~\cite{kavokin03b} for exciton polaritons in a microcavity. The electric field in a photon mode is then given by
\[
 E(\omega)=\sum\limits_{m=1}^{N+1}\frac{C^{(m)}_{\rm phot}\langle C^{(m)}
|F\rangle}{\Omega_m-\omega}\:.
\]
Here the scalar product $\langle C^{(m)} |F\rangle \equiv
\sum_{j=1}^N [C^{(m)}_j]^* F_{j}$ is defined.  Photoluminescence intensity is given by the expression
\begin{equation}\label{PLgen}
 I(\omega)=T\overline{|E(\omega)|^2}=T\sum\limits_{m,m'=1}^{N+1}\frac{C^{(m)*}_{\rm
phot}C^{(m')}_{\rm phot}\langle C^{(m)} |S|C^{(m')}
\rangle}{(\Omega_m^*-\omega)(\Omega_{m'}-\omega)}\:,
 \end{equation}
where the correlation matrix $S$ equals to
  \begin{equation}  \label{S}
  S_{ij}=\langle\!\langle F_{i}^{\vphantom{*}}F_{j}^* \rangle\!\rangle\:,
 \end{equation}
and double angular brackets denote averaging over the random sources realizations.
Factor $T$ is proportional to the mirrors transmission coefficient and links the intensity of the field inside and outside of the cavity. Expression~(\ref{PLgen}) shows that the luminescence intensity has, as a function of frequency $\omega$, poles at the eigenfrequencies $\Omega_m^*$ and $\Omega_{m'}$, and is proportional to the photonic fraction $C^{(m)*}_{\rm phot}C^{(m')}_{\rm
phot}$, as well as to the random sources correlation matrix $S$. The calculation of quantities $S_{ij}$ is only possible by using the microscopic approach, which is developed in the following Sec.~\ref{sec:keldysh}. It is shown below that in the relevant case of exciton quantization as a whole and for the not very closely positioned quantum dots matrix $S$ is diagonal, $S_{ij}\propto \delta_{ij}$.
 
We note that the formulae~(\ref{PLgen}) for the photoluminescence spectrum can be simplified in the strong coupling regime where the imaginary parts of eigenfrequencies $\Gamma_m\equiv \Im \Omega_m$ are negligible as compared with the distances between the real parts, i.e.
\begin{equation}\label{SC}
 |\Re\Omega_m-\Re\Omega_{m'}|\gg 
\Gamma_m,\Gamma_{m'}, \quad (m\ne m').
 \end{equation}
Thus, in the sum in Eq.~(\ref{PLgen}) one can neglect the terms with $m\ne m'$, which leads to the expression 
\begin{equation}\label{PLclass}
 I(\omega)=T\sum\limits_{m=1}^{N+1}|C^{(m)}_{\rm
phot}|^2 \frac{\Gamma_m}{(\omega-\Re \Omega_m)^2+\Gamma_{m}^2}\cdot
   \frac{\langle C^{(m)} |S|C^{(m)} \rangle}{\Gamma_{m}}\:,
\end{equation}
generalizing the results of Ref.~\cite{savona96} to the case of a quantum dot system placed in a zero-dimensional microcavity. Condition (\ref{SC}) means that the polaritons are well defined and, therefore, the structure of terms in Eq.~(\ref{PLclass}) can be interpreted in a following way: first two factors describe photon emission spectrum from $m$th polariton state (product of a photonic fraction in this state by a Lorentzian describing the density of states of a given polariton mode), and a third factor describes a steady population of this state, being proportional to the ratio of the polariton generation rate $\langle C^{(m)} |S|C^{(m)}\rangle$ and polariton damping $\Gamma_{m}$.

\subsection{Green's function method}\label{sec:keldysh}
Microscopic description of photoluminescence of quantum dot arrays in microcavities is possible by using the L.V. Keldysh diagram technique. In order to shorten the notation we neglect the tunneling coupling between quantum dots, and in the end of the present section we comment on exciton tunneling influence.

\begin{figure}
\includegraphics[width=0.45\textwidth]{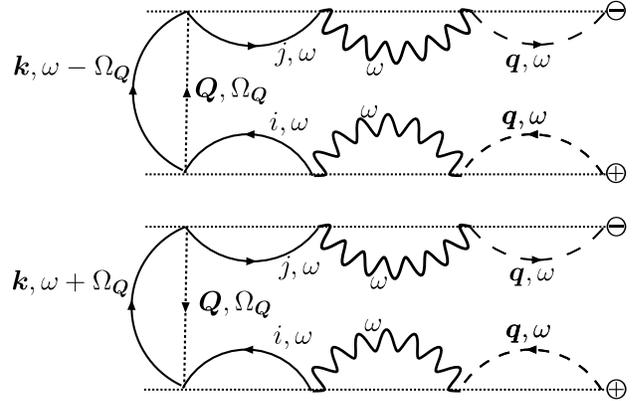}
 \caption{
Diagram representation of Eq.~\eqref{dmp:k}. Dashed lines denote photon Green's function outside of the cavity, wavy lines its Green's function inside the cavity, solid lines are exciton Green's functions in quantum dots calculated without interaction with light, dotted lines are phonon's Green's functions.
}\label{fig:diag}
\end{figure}

One can readily check that the emission intensity is connected with the Green's function of the photon outside of the cavity $D_{\bm q, \omega}^{-+}$ in a following way
\begin{equation}
 \label{intensity:k}
I = -\hbar \omega_{\bm q}\lim_{\gamma\to +0} \left(\frac{\gamma}{\pi} \int D_{\bm q,
\omega}^{-+} d\omega \right),
\end{equation}
where $\gamma$ is a fictitious damping needed to introduce the steady distribution function of photons outside the cavity.

Under low intensity non-resonant pumping the microcavity emission is proportional to the population of the excited exciton states. Therefore, in Dyson equation for Green's functions $D_{\bm q,\omega}^{--}$, $D_{\bm q,\omega}^{++}$ one can totally neglect the pumping, consequently,
\begin{equation}
 \label{phot:greens:k}
D_{\bm q,\omega}^{--} = \frac{1}{\omega - \omega_{\bm q} + \mathrm i \gamma},\:D_{\bm q,\omega}^{++}=-(D_{\bm q,\omega}^{--})^*\:,
\end{equation}
and in equation for $D_{\bm q,
\omega}^{-+}$ it is enough to take into account only linear in excited exciton states populations contributions. As a result, the sought for Green's function can be recast as a sum of two diagrams in Fig.~\ref{fig:diag} (see also \cite{deych:075350}):
\begin{multline}
 \label{dmp:k}
D_{\bm q, \omega}^{-+} \propto |g_{\bm q}|^2 | D_{\bm q, \omega}^{--}|^2 |\mathcal
D_{\omega}^{--}|^2 \sum_{ij,\bm Q} V_iV_j G_{i,\omega}^{++}
G_{j,\omega}^{--}\times \\U_{i\to\bm k}^{\bm Q} U_{\bm k \to j}^{\bm Q} \left[ G_{\bm k,
\omega-\Omega_{\bm Q}}^{-+} N_{\bm Q} + G_{\bm k, \omega+\Omega_{\bm Q}}^{-+}
(1+N_{\bm Q}) \right].
\end{multline}
Here the quantities $g_{\bm q}$ describe the link between the cavity mode and a continuum of photon states outside of the cavity, $\mathcal D^{--}$ is Green's function of the photon mode inside the cavity, calculated with the allowance for the light-matter interaction, $G_{i,\omega}^{--}$ ($G_{i,\omega}^{++}$) are Green's functions of excitons in $i$th quantum dot, $G_{\bm k, \omega}^{-+}$ is Green's function of excitons in the excited states which are numbered by a subscript $\bm k$ (in a general case it runs through the values belonging to both discrete and continuous spectrum). Matrix elements $U_{i\to \bm k}^{\bm Q}$ ($U_{\bm k \to i}^{\bm Q}$) in Eq.~\eqref{dmp:k}
describe the transitions from the state $i$ to the state $\bm k$ (and vice versa),
accompanied by an emission or absorption of a phonon with the wavevector $\bm Q$ and frequency $\Omega_{\bm Q}$, $N_{\bm Q}$ is the phonon distribution function. Phonon damping in Eq.~\eqref{dmp:k}  is neglected.

Exciton Green's functions entering Eq.~\eqref{dmp:k} and calculated in the lowest in the pumping power approximation have form
\begin{equation}
 \label{exc:greens:k}
G_{i,\omega}^{--} = \frac{1}{\omega - \omega_i +\mathrm i \gamma_{i}}, \quad
G_{i,\omega}^{++} = -(G_{i,\omega}^{--})^*,
\end{equation}
\[
 G_{\bm k, \omega}^{-+} = -2\pi f_{\bm k} \delta(\omega - \varepsilon_{\bm k}),
\]
where $f_{\bm k}$ is the distribution function in the excited states formed as a result of the non-resonant pumping, and $ \varepsilon_{\bm k}$ is the spectrum of the excited states. A simple shape of Green's functions is related with the fact that the exciton population in the state $i$ at weak pumping is negligible, the damping of the excited exciton states is neglected also.

One has to determine photon Green's function inside the microcavity self-consistently, with allowance for both light-matter interaction, described by the Hamiltonian~\eqref{hamiltonian}, and for the damping of the photon mode caused by photon tunnelling through the mirrors. As it has been done above, we disregard the populations of the excited exciton states and of the photon mode. By solving the corresponding Dyson equations we have
\begin{align}
&\mathcal D_{\omega}^{--} = \frac{1}{\omega - \omega_{\rm phot} +\mathrm i
\gamma_{\rm phot}(\omega) -  \sum_j V_j^2(\omega - \omega_j +
\mathrm i \gamma_i)^{-1}}, \nonumber \\ 
&\mathcal D_{\omega}^{++} = -(\mathcal
D_{\omega}^{--})^*. \label{phot0:greens:k}
\end{align}
Here $\gamma_{\rm phot}(\omega) = \sum_{\bm q} |g_{\bm q}|^2 \delta (\omega -
\omega_{\bm q})$ is the photon mode damping caused by the escape of photon outside of the microcavity, small renormalization of the mode frequency due to the coupling with an environment is neglected.

Substituting Eq.~\eqref{dmp:k} into Eq.~\eqref{intensity:k}, integrating over frequency and passing to the limit $\gamma\to 0$ one can arrive to Eq.~\eqref{PLgen}. At the same time factor $T$ in Eq.~\eqref{PLgen} turns out to be proportional to $\hbar\omega_{\bm q} |g_{\bm q}|^2$. Elements of the random sources correlation matrix $S$
[Eq.~\eqref{S}] up to a common factor equal to
\begin{multline}
 \label{S:micro}
S_{ij}(\omega)\propto \sum_{\bm k,\bm Q} U_{i\to\bm k}^{\bm Q} U_{\bm k \to j}^{\bm
Q} f_{\bm k} [N_{\bm Q} \delta (\omega -\omega_{\bm Q} - \omega_{\bm k}) +\\ (1+N_{\bm
Q}) \delta (\omega +\omega_{\bm Q} - \omega_{\bm k}) ].
\end{multline}
Therefore, according to its microscopic meaning, $S_{ij}(\omega)$ is the exciton generation matrix (i.e. the rate of its density matrix change) in quantum dots. It is worth noting that here we do not issue the challenge of the precise quantitative description of the carrier relaxation in the quantum dot arrays. It is enough to assume that expression~\eqref{S:micro} describes the last phonon-assisted transition in a quantum dot before the formation of an exciton in the ground state~\cite{bastard1990}.

Inasmuch as the expression $\langle C^{(m)} |S|C^{(m')} \rangle$, determining in accordance with Eq.~\eqref{PLgen} photoluminescence spectrum, constitutes a quadratic form with respect to the Hopfield coefficients $C^{(m)}_i$, and Eq.~\eqref{S:micro} is bilinear  with respect to $U_{i\to\bm k}^{\bm Q}$, $U_{\bm k \to j}^{\bm
Q}$,  formulae \eqref{S:micro} for the generation matrix $S_{ij}$ keeps its shape in an arbitrary basis of exciton states $i$, $j$. Exciton states energies do not enter into the expression for $S_{ij}$ at all, therefore, it can be shown that Eq.~\eqref{S:micro} holds with the allowance for the tunneling coupling between quantum dots as well.

Let us compare diagonal and off-diagonal elements of the matrix $S$.  In the relevant case of the exciton localization as a whole the typical distances between the centers of quantum dots $d$ exceed exciton Bohr radius. One can check that in these conditions $S_{ij} \ll S_{ii}$ ($i\ne j$) provided the wavelength of the phonon which causes the transitions from the excited state $\bm k$ to the states $i$ and $j$ is smaller than $d$. Estimations show that for the acoustic phonon energy $\hbar \omega_{\bm Q}\sim 10$ meV (which corresponds to the typical energy gaps in quantum dots) phonon wavelength amounts to the order of 2 nm, i.e. it is significantly smaller than the typical distances between centres of the neighboring quantum dots. It means that the off-diagonal elements of matrix $S_{ij}$ Eq.~\eqref{S:micro} are small as compared to the diagonal ones, thus
 \begin{equation}\label{S:independent}
  S_{ij}\approx S_i \delta_{ij},
 \end{equation}
and the exciton generation in different quantum dots can be considered as independent.

\section{Nonresonant quantum dots influence on the photoluminescence spectra 
}\label{sec:Gen}
The photoluminescence spectrum  \eqref{PLgen} is determined by the structure of exciton-polariton states and the rate of polariton generation in these states. In the present section we analyze how the emission spectra of the cavity with a single quantum dot in resonance with photon mode is affected by the presence of other, non-resonant, quantum dots. Here and in the following sections we take the exciton-photon coupling constants identical for all quantum dots, $V_i \equiv V>0$, and assume the generation matrix $S_{ij}$ to be diagonal.

\begin{figure}
\includegraphics[width=0.45\textwidth]{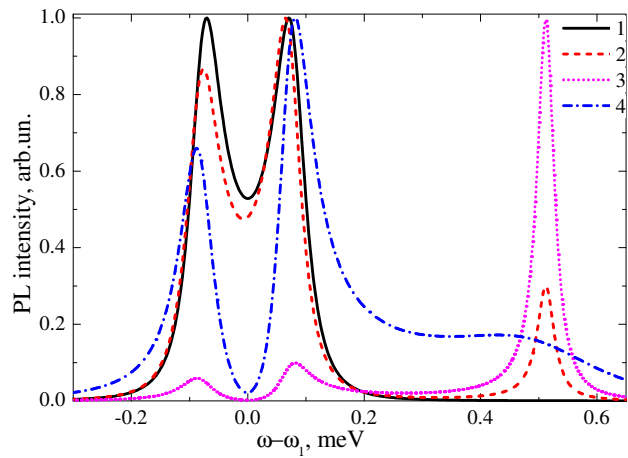}
 \caption{Photoluminescence spectra of the quantum dot system in a microcavity. Solid curve  (1) was calculated for $N=1$, $\omega_{\rm phot}=\omega_1$,  $V=80~\mu$eV, $\gamma_1=17~\mu$eV, $\gamma_{\rm phot}=50~\mu$eV, dashed curve (2) was calculated for $N=2$, $\omega_{\rm phot}=\omega_1$, $\omega_2-\omega_1=0.5~$meV, $\gamma_2=\gamma_1$,  $S_1=S_2$, dotted curve  (3) corresponds to $\gamma_2=\gamma_1$, $S_1=0,S_2\ne 0$ and dash-dotted curve  (4) to $\gamma_{2}=150 \mu$eV, and $S_1=0,S_2\ne 0$ (unspecified parameters are the same as for the first curve). The energy of the photon mode is used as the reference point, the spectra are normalized so that the maximum values coincide. The tunneling between the dots is neglected, $t_{i,j}\equiv 0$.
  }\label{fig:Fabrice}
\end{figure}

The quantum dot photoluminescence in the microcavity was theoretically studied in a recent work~\cite{fabrice08}. 
In the case where only one dot interacts effectively with the cavity mode,  the results of Ref.~\cite{fabrice08} lead  in the limit of weak excitation to Eq. \eqref{PLgen}. An influence of nonresonant quantum dots, with  $|\omega_i-\omega_{\rm
phot}|\gg V$, is taken into account in Ref.~\cite{fabrice08} only phenomenologically: generation and following emission of excitons in these dots 	are reduced to the effective pumping to the photon mode, which, in turn, interacts only with the resonant quantum dots.	
The pumping to the photon mode  can be formally described by adding a random source term $F_{\rm phot}$ 
to the right-hand side of the first of Eqs. \eqref{motioneqs}, the photoluminescence spectrum for $N=1$ is then given by expression
\begin{equation}\label{PLFabrice}
 I(\omega)\propto\frac{S_{\rm phot}|\omega-\omega_1+\rmi\gamma_1|^2+V^2S_1}{|(\omega-\Omega_1)(\omega-\Omega_2)|^2}\,
\end{equation}
where $S_{\rm phot}=\langle\!\langle  F_{\rm phot}^{\vphantom{*}}F_{\rm phot}^* \rangle\!\rangle$. An advantage of our approach is the possibility to take into account explicitly all quantum dots interacting with the cavity mode.  As a result, the quantity $S_{\rm phot}$ can be related to the exciton generation rate,  and the contribution of other dots, lying beyond  the scope of Eq.~\eqref{PLFabrice}, can be found.
  
Fig.~\ref{fig:Fabrice} shows the photoluminescence spectra corresponding to the different mechanisms of exciton generation. The calculation was carried out for the typical values of the quantum microcavity parameters~\cite{hennessy07}, indicated in the caption to  the Figure. Curve 1 was calculated for the case where only a single quantum dot is in the microcavity, so that the spectra demonstrates a doublet  corresponding to the strong-coupling regime. The energies of non-interacting photon and exciton modes are assumed to be identical and chosen as a reference point, the pumping takes place only to the exciton state. 

The  curve 2 demonstrates the photoluminescence spectrum for the system where a second dot, detuned at the value $\omega_{2}-\omega_{\rm phot}=0.5$~meV $\gg V=80$~$\mu$eV from the cavity resonance, was added, the exciton generation rate is the same in both dots, $S_2=S_1$. Tunneling interaction between the dots is neglected. One can see from the comparison of curves  1 and 2,  that in the  frequency range  $|\omega-\omega_{\rm phot}|\ll |\omega_2 - \omega_{\rm phot}|$ the presence of a second quantum dot  leads only to a small  distortion of the spectrum explained by the alteration of polariton eigenfrequencies. Analysis shows that the relative magnitudes of the peaks in this doublet strongly depend on the detuning 
$\omega_{1}-\omega_{\rm phot}$ and on the ratio of the generation rates $S_1/S_2$.  
The rise of the extra peak in the spectra at the exciton resonant frequency in the second dot $\omega_2$ is more important.
   
   This peak at the frequency $\omega_2$ becomes dominant in the spectrum (curve 3) when  the generation of excitons takes place only in the second dot, i.e. $S_1=0,S_2\ne 0$.  Since the second dot  acts also as a photon emitter, the doublet near the frequency $\omega_1$ remains in the spectrum, however, its intensity is smaller as compared to that of the peak at the frequency $\omega_2$. The shape of the doublet is described by Eq.~\eqref{PLFabrice} with $S_1\equiv 0$ and 
\begin{equation}\label{Sphot}
 S_{\rm phot}=S_2\frac{V^2}{|\omega-\omega_2+\rmi\gamma_{2}|^2}\:,
\end{equation} 
 so that $S_{\rm phot}\ll S_2$ when $|\omega-\omega_2+\rmi\gamma_{2}| \gg V$. Note that the depth of the dip between the peaks in the doublet for the curve 3 is considerably larger than for the first two curves. This distinctive feature of pumping ``to the photon mode'' is related to the factor  $|\omega-\omega_1+\rmi\gamma_X|^2$ in the numerator of Eq.~\eqref{PLFabrice}, which is minimal at the frequency  $\omega=\omega_1$.

At last, the curve 4 is obtained for $S_1=0,S_2\ne 0$ and  $\gamma_{2} \gg \gamma_{1}$\:. This calculation models the situation when the generation of excitons takes place, for example, in a continuum of the states (wetting layer) and a direct generation of excitons in the resonant state of first quantum dot is inefficient. Comparing curves  3 and 4 one can see,  that the 
  competition between the emission from resonant and nonresonant exciton states exists at $S_1=0$.

Strong-coupling regime holds for  the parameters used: two polariton frequencies  are $\Re\Omega_{\pm}\approx\omega_1\pm V$ with the decay rates $\Gamma_{\pm}\sim \gamma_{1}+\gamma_{\rm phot} <V$. 
The third ``polariton'' frequency is obviously close to the frequency of the nonresonant quantum dot. The luminescence spectrum is well described by Eq.~\eqref{PLclass}, in this case the  structure of this expression is so that the peak values of the spectra at the frequencies  $\Re\Omega_m$ are proportional to $1/\Gamma_m^2$. The values of PL integrated over the frequency аround each peak are proportional to $1/\Gamma_m$, since the stationary number of the polaritons in the state  $m$ is the ratio of their generation and decay rates.  Therefore the emission from the $m$-th polariton state is more efficient for smaller decay rates $\Gamma_m$.  Consequently, the direct emission from the state 2 is more efficient if the quantity $\gamma_{2}$ is small,
and the growth of  $\gamma_{2}$ leads to the increasing emission from the resonant state $1$, in agreement with Fig.~\ref{fig:Fabrice}.

 Thus, we have demonstrated that the luminescence spectrum of the quantum dots in a strong-coupling regime 
   considerably changes when additional, non-resonant exciton states exist in the system.
   The features at the frequency $\omega_2$ rise in the spectrum against a background of the resonant doublet near the frequency $\omega_{1}$, and the ratio of the photoluminescence intensities at the frequencies $\omega_{1}$
and  $\omega_2$ strongly depends on the parameters of the problem.

%----------------------------------------------------------------------------------------------------------------------------
\section{Role of inhomogeneous broadening }

Now  we proceed to the case where several quantum dots with the energies close to the photon mode energy are present in the microcavity. In  the given section we analyze the role of the  exciton resonant frequencies dispersion due to the inhomogeneous broadening in the ensemble, neglecting the tunneling coupling between the dots, the influence of the tunneling is studied in the next section. 

It is instructive for further consideration to recall the results  \cite{houdre96} for identical quantum dots without tunneling, i.e. $\omega_i\equiv\omega_X$, $\gamma_i\equiv\gamma_X$ and 
$t_{i,j}\equiv 0$. One can easily show that the luminescence spectrum for such a system is still described by Eq.~\eqref{PLFabrice} with $S_2=0$. There are two mixed (polariton) states with the frequencies
\begin{equation}\label{OmegaPM}\begin{split}
 \omega_{\pm}=&\frac{\omega_X+\omega_{\rm phot}-\rmi(\gamma_X+\gamma_{\rm phot})}{2}\pm\\
 &\sqrt{\left[\frac{\omega_X-\omega_{\rm phot}-\rmi(\gamma_X-\gamma_{\rm phot})}{2}\right]^2+NV^2}\:
 \end{split}
\end{equation}
in this case. All other  $N-1$ exciton states do not interact with the photon mode and are not manifested in the luminescence spectrum. It is evident from Eq.~\eqref{OmegaPM} that  the splitting between the energies of polariton states grows proportionally to  $\sqrt{N}$.
  
  This result can be interpreted by analyzing the eigenvectors  $P^{(i)}=\left[P_1^{(i)}\ldots
P_N^{(i)}\right]$ for the ensemble of  not interacting with light quantum dots, determined by the Hamiltonian \eqref{hamiltonian} with
$V=0$.  The energy spectrum is $N$-fold degenerate  for $\omega_i\equiv\omega_X$ and  $t_{i,j}\equiv 0$, so all the  eigenfrequencies  are equal to $\omega_X$, and the eigenvectors $P^{(i)}$ can be chosen as follows:
\begin{equation}\label{SR}
 P^{(1)}_j=\frac{1}{\sqrt{N}},\:\: j=1\ldots N;\quad \sum\limits_{j=1}^N
P^{(i)}_{j}=0,\:\: i\ge 2\:.
\end{equation}
It follows from the first of Eqs.~\eqref{motioneqs}, that for $V_i\equiv V$ the strength of the photon interaction with quantum dots is characterized by the quantity $\sum_{j=1}^N P_{j}$, i.e. it is at maximum when excitons in different dots oscillate in the same phase. Therefore, the light interacts only with one of the modes  \eqref{SR}, the symmetrical
  {\it superradiant} mode $P^{(1)}$  \cite{scheibner2007,Dicke54a,houdre96,kavokin03b,khitrova2007,ivchenko1994,Ikawa_Cho2002}. This interaction leads to the formation of the exciton-polaritons with eigenfrequencies \eqref{OmegaPM}, and all  other $N-1$ exciton states are optically inactive.
  
\begin{figure}
\includegraphics[width=0.4\textwidth]{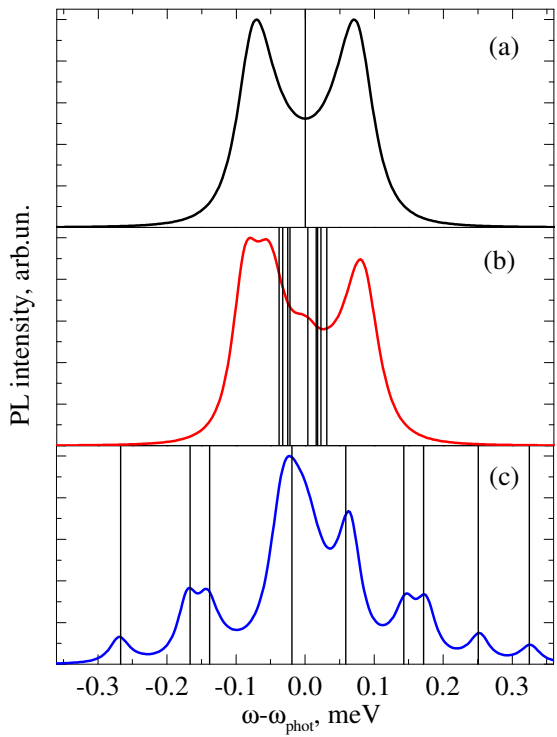}
 \caption{Photoluminescence spectra for  $N=9$ QDs with inhomogeneous broadening taken into account. Panels~(a), (b), (c) were calculated for the cavity mode frequency  $\omega_{\rm phot}=\bar\omega$ ($\bar\omega$ is the average value of the exciton resonant frequencies) and inhomogeneous broadening $\sigma_\omega=0$,  $\sigma_\omega=40~\mu$eV,  $\sigma_\omega=~200~\mu$eV, respectively. Vertical lines indicate the positions of exciton energies when light-matter interaction is neglected. Other parameters are as  follows: $\sqrt{N}V=80~\mu$eV, $\gamma_{\rm phot}=50~\mu$eV, $\gamma_{i}\equiv 17~\mu$eV. The pump is assumed to be identical for all QDs. Vertical scale is the same for all panels, the spectra are normalized so that their maximum values coincide.
 }\label{fig:inhom_pl}
\end{figure}

The light-matter interaction can be conveniently described in the basis of the states \eqref{SR}  with allowance for the dispersion of the quantum dot resonant frequencies. It is evident that the polariton eigenstates are determined by the competition of the exciton interaction with the cavity mode and the inhomogeneous broadening. The first effect keeps the states in the form \eqref{SR},
and the second one splits the symmetrical state to the individual localized exciton modes. The two peak structure of the luminescence spectrum should be destroyed when the dispersion of the resonant frequencies is comparable to $\sqrt{N}V$ rather than to $V$.

This competition is illustrated by the Fig.~\ref{fig:inhom_pl}, where three photoluminescence spectra calculated for $N=9$ different quantum dots with tunneling neglected are shown.  The panel   (a) corresponds to zero inhomogeneous broadening, so the spectrum consists of two peaks splitted by  $2\sqrt{N}V$. Nine exciton frequencies were randomly chosen from the Gaussian  distribution with the dispersion $\sigma_\omega = 40$~$\mu$eV for the panel  (b). The dispersion of the frequencies is sufficiently large, it is greater than the light-matter coupling constant for the individual quantum dot, but smaller, than $\sqrt{N}V$, $\sigma_\omega = 40$~$\mu$eV $ > V \approx 27$~$\mu$eV, however, the two peak structure of the spectrum is still well resolved, with the depth of the dip between the peaks being of the same order as for the panel (a). Calculation demonstrates that the anticrossing of the modes happens when the detuning of the microcavity is changed. For the last panel  (c) of the Fig.~\ref{fig:inhom_pl} 
inhomogeneous broadening is large, $\sigma_\omega = 200$~$\mu$eV $> \sqrt{N}V = 80$~$\mu$eV. The separate peaks corresponding to individual dots in a weak-coupling regime with the cavity mode are resolved here, the alteration of the detuning does not lead to an anticrossing.

Note, that the transition between the weak- and strong-coupling regimes in the quantum dot ensembles with large homogeneous broadening (decay rate) $\gamma_{\rm phot}, \gamma_i > \sigma_\omega$ 
is also determined by the ratio of the decay parameters and the quantity $\sqrt{N}V$~\cite{houdre96}.  Consequently,
if the characteristic broadenings of the photon mode and the quantum dot spectra are smaller than  $\sqrt{N}V$, the microcavity is in the strong-coupling regime and the emission spectrum has a doublet structure. Complex multi-peak spectrum corresponding to the isolated dots in a weak coupling regime takes place in the opposite case.

\section{The role of the exciton tunneling between the quantum dots}\label{sec:tunnel}
In this section the influence of the exciton tunneling between the quantum dots on the photoluminescence spectra is analyzed.
In the general case the tunneling coupling between the dots can obviously lead to the complete alteration of the eigenmodes Eq.~\eqref{SR},  and to the complex, multi-peak photoluminescence spectrum, similar to that shown on Fig.~\ref{fig:inhom_pl}(c). 
However, it turns out that the shape of the spectrum strongly depends on the specific structure of the matrix $t_{i,j}$, i.e.
on the character of the tunneling couplings between the dots. Two possible limiting cases are studied below:
random tunneling only between  neighbor dots and random tunneling between all pairs of quantum dots. Note that the sign of the tunneling contants  $t_{i,j}$ is determined by the form of the wavefunctions corresponding to the exciton states in the dots  $i$ and $j$. If these states are identical and of lowest energy then $t_{ij}<0$, in the general case the sign of $t_{i,j}$ can be arbitrary.

 The shape of the microcavity  luminescence spectrum is qualitatively determined by the distortion degree of the superradiant (fully symmetric) mode~\eqref{SR}, as has been ascertained in the previous section. 
 For the following analysis it is convenient to choose the eigenvectors $P^{(i)}$ in \eqref{SR}  with $i>2$,  corresponding to optically inactive polariton states, as 
\begin{gather}
%  P^{(1)}_j=\frac1{\sqrt{N}}, \quad
 P^{(i)}_j=\frac1{\sqrt{N}}(\delta_{j,1} - \delta_{j,i}),\:\: i=2\ldots N, \:\: j=1\ldots N,\label{srnew}
 \end{gather}
and introduce the tunneling Hamiltonian matrix elements  $\langle i |t |j\rangle$ in the basis ~\eqref{srnew}, $\langle i |t |j\rangle =\sum_{l,m=1}^NP^{(i)}_lt_{l,m}P^{(j)}_m$.  One can show that if all the random quantities $t_{i,j}=t_{j,i}$ 
are independent and identically distributed, the averaged over disorder realizations values of  $\langle i |t |j\rangle$
are equal to 
\begin{align}\label{t_rand}
\overline{ \langle 1 |t |1\rangle}&=(N-1)\bar
t, \quad \overline{\langle 1|t |j\rangle}=0\:,
\end{align}
and all the other matrix elements are in order of  $\bar t$ with root-mean-square deviations being equal to $\sigma_t$ up to the factors of the order of unity. Here $\bar t$ and $\sigma_t$ are the mean value and the dispersion of the random quantity $t_{i,j}$, respectively. 
Such a situation is realized where the  quantum dot array is sufficiently dense, so that each dot is coupled with all (or at least with most part of the other) dots in the ensemble.

\begin{figure}
\includegraphics[width=0.4\textwidth]{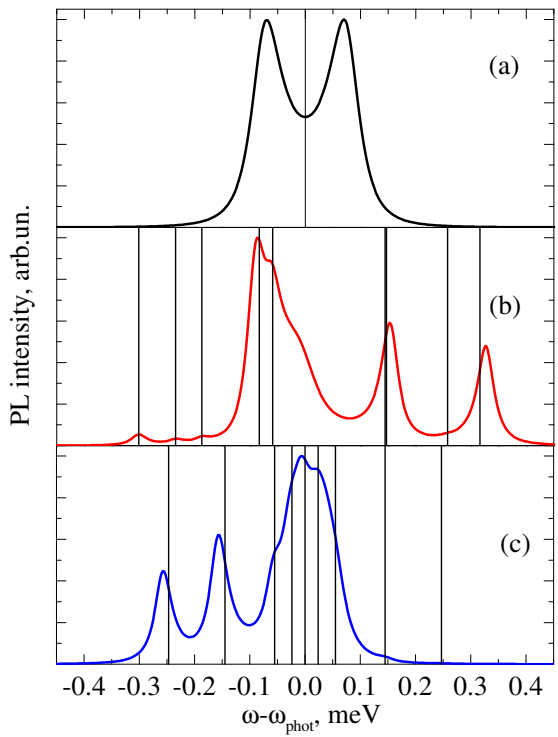}
 \caption{Photoluminescence spectra for $N=9$ QDs, tunneling between the dots is taken into account. Curves  (a) and (b) are calculated for tunneling possible between each pair of QDs,
 $\bar t=-80~\mu$eV и $\bar t=0$,  respectively, and curve (с) is calculated for tunneling only between nearest neighbors, $\bar t=-80~\mu$eV. Tunneling constants dispersion is $\sigma_t=80~\mu$eV, other parameters are the same as for Fig.~\ref{fig:inhom_pl}\:. Vertical lines indicate the positions of exciton energies when light-matter interaction is neglected. 
   Curve (a) corresponds to $\omega_{\rm phot}=\varepsilon_1$ (so the cavity is tuned to the tunnel-split-off mode frequency); curves  (b) and (c) correspond to $\omega_{\rm phot}=\omega_i$. The dispersion of exciton resonant frequencies $\omega_i$ is neglected. Vertical scale is the same for all panels, the spectra are normalized so that their maximum values coincide.
  }\label{fig:tNN_pl}
\end{figure}

It follows from Eqs.~\eqref{t_rand} that for sufficiently large  $N$ and  $\bar t\ne 0$, the spectrum of the states found with tunneling taken into account consists of a single level with the energy  $\varepsilon_1$, corresponding to the superradiant mode  $P^{(1)}$, split-off by the distance in order of  $N|\bar t|$ from the other spectrum. Since the split-off mode weakly interacts with others, all   $N$ exciton states can still be classified as one, interacting with the cavity mode and  $N-1$ practically optically inactive. Therefore, the spectrum keeps the doublet shape, see Fig~\ref{fig:tNN_pl}(a).  The curve shown there, is calculated for the cavity mode being tuned to the tunneling-split-off  exciton state. Besides, the spectrum also possesses a fine structure in the energy region of the remaining exciton states, not shown on the Figure.
 Note that for the calculation presented the tunneling constants dispersion  is of the same order as the photon coupling constant with the superradiant mode, $\sigma_t = \sqrt{N}V$. When  $N$ increases, the strong-coupling regime becomes more stable because both the effective coupling constant and the distance between the spit-off mode and the other spectrum grow.

The structure of the spectrum dramatically changes at $\bar t=0$ (tunneling coupling constants between the dots being of the different signs) or when the tunneling takes place only between the nearest neighbors (non-zero and random elements in the matrix $t_{i,j}$ are only below and above the diagonal). One can readily show that all the matrix elements $\overline{ \langle i |t |j\rangle}$ are then of the same order and the situation  is similar to the inhomogeneous dispersion of the energies of the exciton in individual quantum dots. The spectra has a doublet structure when  $\sigma_t \lesssim \sqrt{N}V$, in the opposite case the eigenmodes are completely altered, in agreement with Fig.~\ref{fig:tNN_pl}(b) (tunneling constants of different signs, mean value $\bar t=0$), (c) (tunneling in the linear geometry between nearest neighbors).

The results presented  are relevant  for the case, where not very large number of quantum dots is present in the volume of the microcavity,
 $N\sim 1\ldots 10$. When $N$ is large and the dots are chaotically distributed, so that the quantities $t_{i,j}$, $\omega_i$ are random, the structure of the polariton eigenstates becomes the same as for the planar microcavity with inhomogeneous broadening taken into account~\cite{kavokin03b,MalpuechKavokin99}. In case  the dots form an ordered two-dimensional superlattice, the optical properties of the system are determined by the structure of the Bloch states of an exciton~\cite{vorivch2d}.

\section{Conclusions}\label{sec:concl}

A theory of the photoluminescence of quantum dot ensembles placed in zero-dimensional microcavities is developed in this paper.  The random source method, allowing a transparent description of the photoluminescence of such structures as a function of the energy parameters of quantum dots and photon mode, is set forth. The random sources approach is microscopically justified within the framework of the standard diagram technique for non-equilibrium  processes, general  expressions for the correlation matrix of the random sources are obtained.

The influence of the excitation parameters, cavity mode energy position and the energy spectrum of the quantum dots on the emission spectra is studied. The luminescence spectra are analyzed for the quantum microcavity where  one dot is in the strong coupling regime with the   photon mode and energy of all  the other ones are far from the  cavity resonance. In this regime the so-called non-resonant pumping to the photon mode, introduced in Ref.~\cite{fabrice08} for the description of the experiments  \cite{reithmaier04a} is justified both qualitatively and quantitatively.

The structure of the eigenmodes and emission spectra in case where all the dots from ensemble are in resonance with the cavity mode is studied in details. An influence of the quantum dot exciton energies dispersion and of the random tunneling coupling between the dots on the photoluminescence spectra is analyzed. It is shown, that the spectra are determined by the competition of the formation of fully symmetric (superradiant) mode and the inhomogeneous broadening in the ensemble. The superradiant mode formation is highly sensitive to the geometry of the tunneling-coupled quantum dot array,  particularly, it is strongly suppressed when the tunneling is possible only between the nearest neighbors. On the other hand, in the dense  enough  QD ensembles   so, that each dot is effectively coupled with many others, the superradiant mode becomes more stable.

The authors are grateful to N.A. Gippius,  F.P. Laussy, E. del Valle and P. Senellart for their interest to the work and useful discussions. The work was supported by RBFR and the programs of RAS. M.M.G. and  A.N.P. acknowledge the support of the ``Dynasty'' Foundation -- ICFPM.

\end{document}